\pdfoutput=1
\documentclass{article}

\usepackage{arxiv}

\usepackage[utf8]{inputenc}
\usepackage[T1]{fontenc}

\usepackage{microtype}
\usepackage{xcolor}
\usepackage{hyperref}
\usepackage{xurl}
\usepackage{booktabs}
\usepackage{amsmath,amssymb,amsfonts}
\usepackage{graphicx}
\usepackage{enumitem}
\usepackage{placeins}
\usepackage{float}
\usepackage{needspace}
\usepackage{tikz}
\usepackage[numbers,sort&compress]{natbib}
\usepackage{doi}
\usetikzlibrary{arrows.meta,positioning,calc}

\graphicspath{{figures/}}

\newcommand{\figgradientwidth}{\columnwidth}
\newcommand{\figsystemswidth}{\columnwidth}

\title{Declared but Unaccountable:\\ Measuring the Seller-Identity Gap \\ in Advertising Supply-Chain Transparency}

\author{
Tommaso Grotto \\
Kopjra Srl, Bologna, Italy \\
\texttt{tommaso.grotto@kopjra.com} \
\And
Alessandro Cantelli-Forti \\
CNIT, RaSS Laboratory \\
Pisa, Italy \\
\texttt{alessandro.cantelli.forti@cnit.it} \
\And
Alessio Merlo \\
CASD, School for Advanced Defense Studies \\
Rome, Italy \\
\texttt{alessio.merlo@unicasd.it}
}

\date{}
\renewcommand{\shorttitle}{Declared but Unaccountable}

\hypersetup{
  hidelinks,
  pdftitle={Declared but Unaccountable: Measuring the Seller-Identity Gap in Advertising Supply-Chain Transparency},
  pdfsubject={Seller-account accountability in programmatic advertising},
  pdfauthor={Tommaso Grotto, Alessandro Cantelli-Forti, Alessio Merlo},
  pdfkeywords={advertising transparency, ads.txt, app-ads.txt, sellers.json, seller-identity gap, supply-chain accountability, Web measurement}
}

\begin{document}
\maketitle

\begin{abstract}
Public advertising transparency relies on a minimum accountability condition: an account authorized by a publisher in \texttt{ads.txt} or \texttt{app-ads.txt} should appear in the applicable \texttt{sellers.json} registry. When an available registry omits the declared seller ID, the authorization remains visible and machine-readable, but its public verification path breaks. We call this failure the \emph{seller-identity gap}.

We measure the gap using 233.4 million normalized advertising-transparency records and an absent-only, coverage-aware evidence rule. An eligible declaration enters the numerator only when the applicable registry is available and does not contain its normalized seller-ID key; unavailable, malformed, confidential, and otherwise non-evaluable states remain separate. The resulting row-level floor is 24.25\% in a high-popularity, low-removal reference population and 30.71\% among publishers associated with at least 100 copyright-removal requests, a 6.45 percentage-point increase. The gap then remains near one third at stricter cumulative thresholds. Standardizing the populations to common weights over 1,469 shared ad systems reduces the primary contrast to 4.65 points: about 72\% remains within systems, while 28\% forms a composition-associated residual that also includes restriction to shared system support.

We connect these population estimates to recurring seller-account coordinates, byte-identical declaration configurations, and provenance-bearing case packets. We frame this declaration-centered investigative use of public advertising records as \emph{Ad Transparency Intelligence}: it reconstructs declared supply-chain relationships, identifies where public verification stops, and prioritizes evidence for corroboration without equating correlation with attribution. The result is a reproducible path from population-scale measurement to reviewable declaration evidence. Advertising transparency supports accountability and investigation only when visible authorizations remain publicly verifiable.
\end{abstract}


\section{Introduction}
\label{sec:introduction}
Online advertising finances much of the open Web, including sites associated with piracy, misinformation, illegal streaming, scams, and other forms of online abuse~\cite{batikas2019followmoney,rafique2016freestreaming,sheaib2025unmasking,euipo2026iprads,papadogiannakis2023whofunds,dca2021breakingbads,dca2022ransomware}. For an advertiser, media buyer, regulator, or investigator, following the money begins with a concrete question: \textsl{which accounts and intermediaries are positioned to monetize this publisher?} A single impression may pass through publishers, supply-side platforms, exchanges, and resellers before reaching a buyer. Reconstructing declared relationships is therefore a prerequisite to accountability, even when execution and payment remain unobservable~\cite{ana2023mfa,isba2020supplychain,kroll2023qualitative}.

\begin{figure}[t]\centering
\resizebox{\columnwidth}{!}{
\begingroup
\newcommand{\flab}[1]{{\fontsize{6.5}{7}\selectfont\ttfamily\textcolor{black!50}{#1}}}
\begin{tikzpicture}[
  font=\small,
  box/.style={draw, rounded corners=1.6pt, line width=0.5pt, align=center,
              minimum height=8.5mm, inner xsep=2.4mm, inner ysep=1.4mm},
  linklab/.style={font=\fontsize{6.5}{7}\selectfont, text=black!50},
  link/.style={-{Latex[length=1.6mm]}, line width=0.7pt},
]
\definecolor{fail}{HTML}{B3402F}
\node[box] (pub) {Publisher\\[-0.5pt]\flab{ads.txt / app-ads.txt}};
\node[box, fill=black!4, right=8mm of pub] (adsys) {Ad system\\[-0.5pt]\flab{sellers.json}};
\node[box, fill=black!8, right=10.5mm of adsys] (record) {Public seller record\\[-0.5pt]\flab{seller ID present}};
\draw[link] (pub) -- (adsys);
\draw[line width=0.7pt] (adsys.east) -- ++(2.8mm,0) coordinate (bk);
\draw[fail, line width=1.1pt, line cap=round] ($(bk)+(0.7mm, 1.6mm)$) -- ($(bk)+(3.5mm,-1.6mm)$);
\draw[fail, line width=1.1pt, line cap=round] ($(bk)+(0.7mm,-1.6mm)$) -- ($(bk)+(3.5mm, 1.6mm)$);
\draw[link, dashed] ($(bk)+(4.2mm,0)$) -- (record.west);
\node[linklab] at ($(pub.north east)!0.5!(adsys.north west)+(0,2.0mm)$) {declares seller ID};
\node[linklab] at ($(adsys.north east)!0.5!(record.north west)+(0,2.0mm)$) {registry lists seller ID};
\path (pub.south) -- (record.south) coordinate[midway] (bot);
\node[font=\fontsize{7}{8}\selectfont\bfseries, text=fail, below=2.6mm of bot]
  {24.3\% reference $\rightarrow$ 30.7\% $\geq100$ removals};
\end{tikzpicture}%
\endgroup}
\caption{The public seller-account chain. A publisher declares an account in \texttt{ads.txt} or \texttt{app-ads.txt}; the ad system is expected to list the same seller ID in \texttt{sellers.json}. We measure failure only when the applicable registry is available.}\label{fig:chain}
\end{figure}
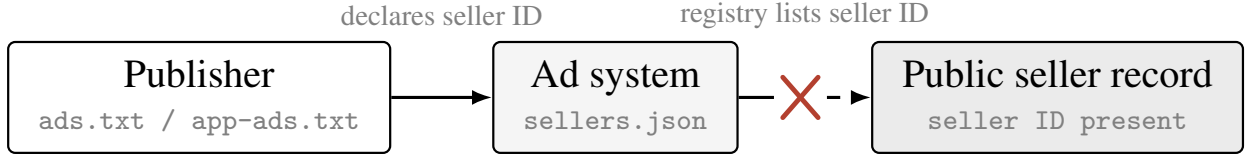

Publishers list authorized seller accounts in \texttt{ads.txt}, or \texttt{app-ads.txt} for applications; ad systems publish reciprocal account records in \texttt{sellers.json}~\cite{iab_adstxt,iab_appadstxt,iab_sellersjson}. Together, these files create a public seller-account chain (Figure~\ref{fig:chain}).
A declaration may remain visible and syntactically valid even when its normalized seller-ID key is absent from the applicable registry. We call this dangling authorization reference the \emph{seller-identity gap}. It breaks the minimum public account-presence property but does not prove fraud, infringement, contractual invalidity, or the absence of a private account.

{{Consider, for example, a publisher that declares the illustrative seller ID \texttt{74291} at an ad system. The declaration is visible, syntactically valid, and consumable by automated tools. If the ad system's available \texttt{sellers.json} contains no record for \texttt{74291}, however, the public chain stops before the account that the declaration is supposed to identify. Buyers, auditors, exchanges, brand-safety systems, and investigators rely on these artifacts as public authorization metadata. A missing record does not prove fraud, but it removes the mechanism's minimum guarantee: an outside verifier can no longer connect the declared authorization to a publicly recorded seller account.}}

Cross-file reconciliation and missing seller IDs are not new. Academic, standards, and industry work already measures adoption, missing identifiers, relationship inconsistencies, pooling, and ecosystem-wide matching~\cite{bashir2019adstxt,vekaria2024darkinventory,papadogiannakis2025darkside,poilve2020adweb,iabeurope2025adoption,moss2021deepsee,moss2021darkpool,smith2020correlation,ivp5adtechaudit}. We instead ask how much of the complete declared authorization surface fails this public account-presence check under an explicit evidence rule, how the resulting population contrast is structured, and how the underlying evidence can support targeted review.

We analyze 233,416,998 normalized advertising-transparency records. All eligible publisher declaration rows remain in the denominator; registry records supply account-presence evidence rather than additional denominator rows. Only an absent normalized seller-ID key in an applicable available registry enters the numerator. Unavailable, malformed, and otherwise non-evaluable registries remain separate; evidence-backed aliases retain their provenance, and confidential records count as account-present without revealing public identity. We compare a high-popularity, low-removal reference with cumulative populations associated with at least 100, 1,000, and 10,000 Google copyright-removal requests~\cite{lepochat2019tranco,google_transparency}. These thresholds indicate enforcement attention, not culpability.

{{The result is both substantial and easy to state: the gap affects almost one in four declaration paths in the reference population and almost one in three at the primary enforcement-linked threshold. The coverage-robust floor rises from 24.25\% to 30.71\% (a 6.45-percentage-point increase), then plateaus near one third. Standardization to common weights over 1,469 shared ad systems reduces the contrast to 4.65 points. Approximately 72\% of the observed difference therefore remains within systems, while the remaining 28\% is composition-associated and includes the small effect of restricting standardization to shared systems. The decomposition localizes the contrast without identifying its cause.}}

{{Beyond measurement, the same public records provide an investigative surface. We use \emph{Ad Transparency Intelligence} to describe this declaration-centered, ecosystem-level use of advertising-transparency artifacts: reconstructing declared supply-chain relationships, identifying where public verification fails, and prioritizing evidence-bearing review. A resolved coordinate can identify a seller account and the intermediary at which further verification can proceed; an absent account interrupts that public path. This use supports correlation and review prioritization, but does not by itself establish ownership, intent, contractual status, or economic benefit.}}

Accordingly, we ask:
\begin{description}[leftmargin=0pt,labelindent=0pt,style=nextline,font=\normalfont\bfseries]
\item[RQ1: How much of the declared advertising surface fails the public seller-account check?] We estimate the coverage-robust row-level floor and compare the reference with cumulative enforcement-linked populations.
\item[RQ2: How is the enforcement-linked difference structured?] {{We test observed popularity support and separate the contrast that remains under common ad-system weights from the residual associated with ad-system composition.}}
\end{description}

This work makes three contributions:
\begin{itemize}[leftmargin=2.4em,nosep]
\item[\textbf{C1}] We formalize public seller-account presence as a distributed referential-integrity property and operationalize an absent-only, provenance-preserving estimand.
\item[\textbf{C2}] {{We measure a substantial reference floor and a 6.45-point enforcement-linked step, then localize that contrast through a common-standard decomposition without causal attribution.}}
\item[\textbf{C3}] {{We translate population-scale measurement into declaration-native investigative surfaces based on recurring seller-account coordinates, byte-identical declaration configurations, and provenance-bearing case packets.}}
\end{itemize}

{{Identifier reuse, declaration clustering, and advertising intelligence are established practices~\cite{starov2018dashboard,papadogiannakis2022publisherids,adalytics2021network,vines2017adint}. Our contribution is to make their declaration-centered use more controlled and verifiable by preserving the evidence path from population estimates to the public records that produced them. The remainder defines the property and evidence model; positions the work against its nearest neighbors; reports the data, estimand, RQ1, and RQ2; develops Ad Transparency Intelligence through reviewable declaration-native evidence; and closes with implications, limitations, conclusions, ethics, and open science.}}

\section{Public Advertising Accountability}
\label{sec:accountability}
{{This section defines the public mechanism, the security property it is expected to provide, and the evidence available to an outside verifier. The measured property is verifiable seller-account presence: whether a publisher's declared authorization reaches the public account record promised by the applicable registry. Legal identity, private authorization, and auction execution require additional evidence.}}

\subsection{The Public Authorization-to-Account Chain}\label{sec:public-chain}
An \texttt{ads.txt} row contains an ad-system domain, seller identifier, and relationship type, typically \texttt{DIRECT} or \texttt{RESELLER}; \texttt{app-ads.txt} is analogous for applications. The reciprocal \texttt{sellers.json} maps seller IDs to account records or marks identity information confidential~\cite{iab_adstxt,iab_appadstxt,iab_sellersjson}. The OpenRTB SupplyChain object describes parties in a bid-request path, but it is a distinct execution-layer object~\cite{iab_openrtb,iab_adcom,iab_vast43}. Reselling is legitimate~\cite {jounce2025reselling}.

Seller identifiers are local to an ad system. The same identifier at two systems does not denote the same account. The namespace-safe coordinate is $(\textit{canonical ad system},\textit{seller ID})$. At this coordinate, the publisher file expresses authorization, and the registry establishes whether a public account record exists. {{A resolved coordinate gives an outside verifier both a seller account and the public intermediary at which corroboration can continue; an absent coordinate interrupts that path.}} We separate registry availability, account presence, identity disclosure, observable execution, and economic outcome. The headline estimand ends at account presence. A confidential record is account-present without providing full public identity disclosure.

The term advertising transparency also denotes platform-side explanations and public ad repositories~\cite{andreou2018adtransparency,mozilla2024fulldisclosure,benzaamia2026dsa,angus2024observability,angus2024adobservatory}. Here it refers only to supply-side declaration and registry files.

\subsection{The Seller-Identity Gap}\label{sec:seller-identity-gap}
For ad system $a$, seller identifier $s$, and applicable registry $R_a$, the minimum invariant is
\begin{equation}
\mathrm{declared}(a,s)\wedge\mathrm{available}(R_a)\Longrightarrow\mathrm{present}(s,R_a).
\end{equation}
The seller-identity gap occurs when $(a,s)$ is declared, $R_a$ is available, and $s$ is absent. If the registry cannot be retrieved or parsed, public evidence is insufficient, and the declaration is not counted as absent. Evidence-backed aliases may locate a registry but cannot create a missing ID match. Fuzzy matching, company-name similarity, and corporate affiliation likewise cannot recover a missing coordinate.

Account resolution, identity disclosure, and entity accountability are successively stronger. Names and domains may be missing, confidential, inconsistent, or associated with a parent company. A failed join can reflect a stale or copied file, error, migration, registry omission, unpublished private account, or fabrication. Public artifacts do not reveal which explanation applies. Account presence is necessary but insufficient for complete seller accountability. {{The gap therefore measures a verifier-visible break in the public chain. It identifies where a declared relationship can no longer be followed through the mechanism's own records, without determining why the break occurred or who is responsible.}}

\subsection{Threat and Evidence Model}\label{sec:threat-evidence-model}
We consider an outside verifier, such as a buyer, publisher, ad system, standards body, researcher, regulator, or investigator. The verifier observes public declarations, registries, endpoints, redirects, timestamps, hashes, retrieval states, and parser outputs. It does not observe private contracts, onboarding or KYC records, internal account status, auction logs, settlement, or payment. Its goal is to determine whether a public authorization terminates at the account record promised by the public mechanism.

The design is fail-closed: missing evidence remains missing evidence. {{A registry that cannot be evaluated does not become evidence of an absent account, and a resolved account does not by itself establish a legal entity, active transaction, or economic outcome. These boundaries preserve the value of the public check without extending its inference beyond the evidence available to the verifier.}} Industry analyses similarly show that public programmatic records provide only part of end-to-end supply-chain accountability~\cite{iabey2015untrustworthy,ana2023mfa,isba2020supplychain}.

\subsection{Enforcement-Linked Populations and Scope}\label{sec:enforcement-scope}
Google copyright-removal requests are used only for population construction~\cite{google_transparency}. Removal volume reflects enforcement attention and may depend on scale, visibility, rights-holder effort, reporting intensity, and duplicated URLs. It is neither ground truth for infringement nor a measure of intermediary intent. The reference is high-popularity and low-removal rather than verified benign. The enforcement-linked populations are cumulative observational strata, not causal treatments. Results are conditional on usable transparency declarations and describe declaration rows, not all candidate hosts, all removal targets, or publisher prevalence.

\section{Related Work}
\label{sec:related-work}
We position this work by the quantity measured, the evidence retained, and the inference supported. The literature established cross-file reconciliation, missing seller identifiers, identifier reuse, and declared-plus-observed advertising analysis. Building on those foundations, we contribute a coverage-aware row-level floor for public seller-account non-resolution, compare it across enforcement-stratified populations, decompose the observed contrast into within-system and composition components, and preserve a path from aggregate estimates to reviewable declaration evidence.

\subsection{Transparency Standards and Cross-File Consistency}
\label{sec:related-transparency}
The IAB transparency stack defines complementary declaration, registry, and execution-path layers~\cite{iab_adstxt,iab_appadstxt,iab_sellersjson,iab_openrtb,iab_adcom,iab_vast43,iab_transparency_center}. Bashir et al. measured adoption, authorization compliance, and evolution of \texttt{ads.txt}~\cite{bashir2019adstxt}. CNIL/LINC recursively mapped seller and intermediary relationships, while IAB Europe reported line-level cross-file matching under partial ad-system coverage~\cite{poilve2020adweb,iabeurope2025adoption}. These studies establish adoption and incompleteness but do not provide our absent-only row-level population contrast.

{{The reported quantities also differ materially. IAB Europe found that approximately 79\% of valid \texttt{ads.txt} lines matched a \texttt{sellers.json} record, while registry collection succeeded for only about 61\% of unique ad-system domains~\cite{iabeurope2025adoption}. DeepSee reported matches for roughly 60\% of \texttt{ads.txt} entries across more than one million publishers and for about 40\% of unique reseller entries~\cite{moss2021deepsee,moss2021darkpool}. Vekaria et al. reported the relevant missing-ID category for 65\% of 251 rank-matched control sites and 83\% of 251 misinformation sites~\cite{vekaria2024darkinventory}. Sheaib et al. reported 169 missing seller IDs and 418 domain mismatches among 587 unique \texttt{DIRECT} entries from 35 illegal-streaming sites~\cite{sheaib2025unmasking}. A public audit based on a May 2026 snapshot covered 86,357 publishers and 33.39 million publisher--system--seller triples; within its 6.43 million \texttt{DIRECT} claims, 30.51\% failed to reconcile~\cite{ivp5adtechaudit}. Its aggregate also includes broken registries, confidential records, and registry-migration states that our resolver separates. These figures are not numerically interchangeable with our floor because they differ in unit, relationship scope, population, registry-availability treatment, alias handling, state aggregation, and denominator.}}

Vekaria et al. are the nearest academic comparison~\cite{vekaria2024darkinventory}. They classify seller IDs declared by publishers but absent from corresponding registries and compare misinformation and control sites. Their endpoint is primarily site-level and embedded in a broader dark-pooling study. One absent row and thousands can therefore have the same site-level outcome. Sheaib et al. cross-check \texttt{DIRECT} declarations from illegal streaming services and report missing IDs and domain mismatches~\cite{sheaib2025unmasking}. Their population is similarly policy relevant, but their scope and denominator differ. Papadogiannakis et al. audit related conformance failures at graph scale, including relationship mismatches, confidentiality, copied registries, and hidden intermediation~\cite{papadogiannakis2025darkside}. They do not report our availability-conditioned all-row floor, enforcement gradient, clustered uncertainty, or common-standard decomposition.

{{The estimand distinction is operational. A site-level indicator asks whether a publisher exposes at least one problematic coordinate. A conditional match rate excludes unavailable evidence from its denominator. Analyses restricted to \texttt{DIRECT} declarations exclude reseller relationships. Our floor instead measures absent seller accounts over the complete declared authorization surface while preserving unavailable, malformed, aliased, confidential, and noncanonical states.}}

\subsection{Abuse-Linked Advertising, Pooling, and Remediation}
\label{sec:related-pooling}
Advertising-funded piracy, misinformation, and illegal streaming have been studied through economic, ecosystem, and follow-the-money analyses~\cite{batikas2019followmoney,rafique2016freestreaming,papadogiannakis2023whofunds,euipo2026iprads,dca2021breakingbads,dca2022ransomware,bozarth2021marketforces,ahmad2024misinformation,usc2013adtransparency}. Industry studies document opacity, waste, and untraceable spend~\cite{iabey2015untrustworthy,ana2023mfa,kroll2023qualitative,isba2020supplychain,googlewhiteops2018_3ve,jounce2025reselling}. Security research separately measures fraudulent exchanges, impression and click fraud, affiliate abuse, ad injection, and malicious advertising~\cite{stonegross2011adexchanges,springborn2013impression,pearce2014zeroaccess,dave2012clickspam,chachra2015affiliate,thomas2015adinjection,zarras2014malvertising,li2012madtracer}. We use this literature to motivate the population comparison, not to label a publisher, coordinate, declaration, or impression as fraudulent.

Pooling concerns reuse of an account across publishers; account absence concerns whether the declared coordinate exists in the applicable available registry. A pooled ID can resolve correctly, while a unique ID can be absent. Vekaria et al. combine declarations with browser-observed delivery while studying dark pooling~\cite{vekaria2024darkinventory}. Papadogiannakis et al. analyze shared identifiers, hidden intermediation, placement laundering, and potential revenue paths~\cite{papadogiannakis2025darkside,papadogiannakis2024adlaundering}; their \emph{Before \& After} study pairs declarations and registries with crawled advertising evidence~\cite{papadogiannakis2025beforeafter}. Declared-plus-observed measurement is therefore established. Vekaria et al. later evaluate multi-stakeholder notification and remediation~\cite{vekaria2024notifications}; we do not conduct notifications or estimate remediation effectiveness.

\subsection{Identifiers and Advertising Intelligence}
\label{sec:related-attribution}
Advertising and analytics identifiers have long been used to connect sites and infer administration or campaign structure~\cite{starov2018dashboard,papadogiannakis2022publisherids,adalytics2021network}. Related studies examine tracking ecosystems, actor relationships, privacy leakage, and identity entanglement~\cite{englehardt2016openwpm,siby2022webgraph,papadogiannakis2021postcookie,sanchezrola2021cookies,urban2020gdpr,venkatadri2018facebook,oh2022cartology,korolova2011privacy}. Our fan-out and byte-identical-file analyses build on this practice and serve only as review surfaces.

{{Vines et al. introduced ADINT in a different investigative setting: purchasing targeted advertisements to obtain information about individuals, including location and use of potentially sensitive applications~\cite{vines2017adint}. We study neither targeted surveillance nor individual users. Our declaration-centered use of advertising intelligence reconstructs publisher--intermediary--account relationships from public supply-chain artifacts, identifies where those relationships cease to be publicly verifiable, and reduces the resulting surface to evidence-bearing review targets. In both settings, advertising artifacts become an intelligence source; the object, evidence, and supported inference differ.}}

{{Infrastructure and registration data can reveal dependencies, but shared hosting, CDNs, certificates, and provider-level features may connect unrelated sites~\cite{han2022infrastructure,hounsel2020infrastructure,cangialosi2016keysharing,alrwais2017bulletproof,lu2021whowas}. Placement laundering further complicates inference from delivery endpoints~\cite{kline2018placement,papadogiannakis2024adlaundering}. We therefore keep the present analysis declaration-native: recurring seller coordinates and byte-identical files preserve exact public relationships, while external technical similarity is not used as an account resolver or ownership signal.}}

\subsection{Declared and Observed Advertising Evidence}
\label{sec:related-execution}
{{Prior work infers advertising relationships and auction behavior from retargeting, header bidding, auction observations, and network flows~\cite{bashir2016retargeting,bashir2018diffusion,cook2020kashf,pachilakis2019waterfalls,aqeel2020headerbidding,musa2022atom,yuan2013rtb,olejnik2014auction,papadopoulos2017product,zeng2022targetingbids}. Vekaria et al. and Papadogiannakis et al. further demonstrate the value of combining public declarations with observed advertising delivery~\cite{vekaria2024darkinventory,papadogiannakis2025beforeafter,papadogiannakis2023whofunds}. These studies establish declared-plus-observed measurement as prior art. They also show that observation introduces a separate sampling and inference problem: visible requests, bids, and transaction stages expose different evidence, while server-side paths may remain hidden. Signed transaction chains and accountable exchanges pursue stronger execution-level properties~\cite{pastor2021adschain,zhong2023addax,mrc2026auction}. We therefore confine this paper to the declared public surface and treat systematic comparison with live RTB traffic as subsequent work requiring its own coverage model and phase-aware identifiers.}}

\subsection{Positioning of This Work}
{{We do not claim priority for cross-file reconciliation, missing-ID detection, identifier reuse, advertising intelligence as a general investigative idea, or declared-plus-observed measurement. We build on those foundations to contribute an explicit public account-presence property; an absent-only, coverage-aware row-level estimand with publisher-clustered uncertainty~\cite{cameron2008bootstrap}; an enforcement-linked population comparison; a common-standard decomposition following the rate-versus-composition logic of Kitagawa~\cite{kitagawa1955components}; and provenance linking population estimates to recurring coordinates, shared declaration configurations, and exact reviewable evidence. The novelty lies in this controlled combination of measurement, comparative localization, and declaration-native traceability.}}

\section{Data and Measurement Design}
\label{sec:measurement}
This section operationalizes the public account-presence property. It keeps acquisition, population, and analytical scales separate and fixes evidence states, analytical unit, denominator, uncertainty, decomposition, and provenance before examining outcomes.

\subsection{Transparency Corpus}
\label{sec:corpus}
The canonical snapshot, frozen for analysis on September 28, 2026, contains 233,416,998 normalized records: 128,964,813 \texttt{ads.txt} rows, 100,558,277 \texttt{app-ads.txt} rows, and 3,893,908 \texttt{sellers.json} rows. The publisher-declaration families describe authorization relationships; seller records provide the reciprocal registry context. The retained observations span March 20--September 21, 2026, with the publisher-declaration families beginning March 26. The surviving campaign ledger covers April 16--September 21: 3,307 artifact-family runs, 51,222,091 application-level fetch attempts, and 24,000,178 present-valid artifact observations. These operational totals describe the surviving run summaries, not a complete census of every historical attempt or the denominator of any analytical result.

The analytical unit is a frozen normalized declaration row, not a fetch attempt. For each declaration, we retain host, artifact family, source hash, source run, parser status, line position, raw and canonical ad-system domains, seller ID, and relationship type. Registry rows retain seller ID, declared name and domain, seller type, confidentiality state, source hash, and parser provenance. Content hashes deduplicate identical bytes and make paper-facing results traceable to frozen inputs. Repeated observations support provenance and reduce reliance on a one-off exposure, but headline estimates remain cross-sectional. We do not describe them as longitudinal change.

\subsection{Populations and Selection Boundary}
\label{sec:populations}
The reference begins with Tranco top-1M~\cite{lepochat2019tranco} and excludes hosts with at least 100 removal requests. It can include hosts with 1--99 requests and hosts for which the source has no evidence. Enforcement-linked populations contain hosts with at least 100, 1,000, and 10,000 requests and are nested.
\begin{table*}[t]
\centering
\small
\caption{Populations and public-registry coverage. The removal populations are cumulative.}
\label{tab:populations}
\begin{tabular}{lrrrr}
\toprule
Population & Candidate hosts & Hosts with declarations & Declaration rows & Registry coverage \\
\midrule
Tranco low-removal reference & 972,214 & 78,686 & 40,384,095 & 96.54\% \\
Google removals $\geq100$ & 485,920 & 8,998 & 6,344,799 & 95.34\% \\
Google removals $\geq1,000$ & 100,836 & 2,227 & 2,626,098 & 95.14\% \\
Google removals $\geq10,000$ & 11,182 & 374 & 691,380 & 95.08\% \\
\bottomrule
\end{tabular}
\end{table*}
The analysis is conditional on a host exposing at least one usable publisher declaration. This occurs for 8.09\% of candidate reference hosts, compared with 1.85\%, 2.21\%, and 3.34\% of candidates in the cumulative removal populations. We therefore make no population-wide claim about all copyright-removal targets. Differing declaration rates are a selection boundary, not an implementation detail.

{{A separately archived June 21 deep-mapping study contains 37,454 removal-linked hosts, 15.3 million grouped declaration leads, and 321,097 seller-account clusters. Grouped leads are not raw normalized rows. Its fan-out statistics provide a dated investigative illustration, not an additional population estimate or an extension of the four RQ1/RQ2 denominators. Exact-content membership and the four illustrative publishers in Section~\ref{sec:audit-surfaces} are checked against the canonical snapshot; their copyright-removal context remains separately dated. We fixed the historical case identities independently of the population results. None of these surfaces changes the primary labels, denominators, common weights, or resolution states.}}

\subsection{Coverage-Aware Resolution}
\label{sec:resolution}
For each declaration row $r$, the resolver preserves the raw ad-system value, canonicalizes the domain, and seeks the applicable registry. Direct registries and evidence-backed aliases are accepted. Alias rules cover verified redirects, rebrands, and special registry hosts; each retains its source and raw endpoint. Registry retrieval terminates as available directly, available through an evidence-backed alias, unavailable, malformed, or not materialized. When an applicable registry is available, account resolution yields resolved, absent, confidential, or noncanonical.

Eligibility requires nonempty host, ad-system, and account-ID fields; ad-system tokens matching the fixed malformed-domain pattern are excluded.\footnote{The analysis filter uses the regular expression \texttt{\detokenize{^[0-9a-f]{12,}}}\allowbreak\texttt{\detokenize{(?:[a-z0-9-]+\.)+[a-z]{2,}}}.} Account presence is tested by exact equality of normalized seller-ID keys within the applicable ad-system namespace, including evidence-backed registry aliases. Keys are trimmed and lowercased on both declaration and registry sides; raw identifiers are retained. This is not fuzzy matching or byte-for-byte equality of raw strings.
\begin{table}[t]
\centering
\scriptsize
\setlength{\tabcolsep}{3.5pt}
\caption{Registry and account states. Alias use is retrieval provenance rather than a mutually exclusive account outcome. Only an absent account in an applicable available registry enters the primary numerator.}
\label{tab:states}
\begin{tabular}{@{}lp{0.49\columnwidth}p{0.28\columnwidth}@{}}
\toprule
Layer & State & Effect on primary numerator \\
\midrule
Registry & available directly or through alias & account evaluated \\
Registry & unavailable, malformed, or not materialized & no \\
Account & resolved & no \\
Account & confidential but present & no \\
Account & absent & yes \\
Account & noncanonical & no \\
\bottomrule
\end{tabular}
\end{table}
Confidential records expose the account while withholding identity and comprise only 0.38--0.69\% of rows across the main populations. Unavailable, malformed, and not-materialized registries do not enter the numerator. This prevents crawler coverage from becoming apparent publisher opacity.

\subsection{Primary Estimand and Uncertainty}
\label{sec:estimand}
Let $D_P$ denote eligible declaration rows in population $P$. Let $\mathrm{loaded}(r)$ indicate an obtained applicable registry and $\mathrm{absent}(r)$ that its normalized seller-ID key does not occur:
\begin{equation}
\mathrm{Gap}(P)=\frac{|\{r\in D_P:\mathrm{loaded}(r)\wedge\mathrm{absent}(r)\}|}{|D_P|}.
\end{equation}
All eligible declarations remain in the denominator. Unavailability can lower the floor but cannot increase its numerator. Registry coverage is reported separately. The estimand is row-weighted and describes the declared authorization surface, not publisher prevalence. Publishers with larger files contribute more rows. We obtain 95\% intervals from 1,000 publisher-level bootstrap samples, retaining each sampled host's complete row set~\cite{cameron2008bootstrap}.

\subsection{Comparative Design and Decomposition}
\label{sec:comparative-design}
Popularity matching compares enforcement-linked and reference publishers within common Tranco rank buckets and is a sensitivity analysis, not a causal adjustment. For population $P$ and system $a$, let $g_{P,a}$ be its floor and $w_{P,a}$ its row share. The common support consists of 1,469 systems present in all four populations. The weighting rule pools declaration rows over these systems and all four population memberships, including the nested cohorts, then normalizes the pooled totals to obtain $w_a^*$:
\begin{equation}
\mathrm{Gap}^*(P)=\sum_a w_a^*g_{P,a}.
\end{equation}
For enforcement-linked $E$ and reference $R$,
\begin{equation}
\Delta_{\mathrm{within}}^*=\sum_a w_a^*(g_{E,a}-g_{R,a}),\qquad
\Delta_{\mathrm{comp}}^*=\Delta_{\mathrm{obs}}-\Delta_{\mathrm{within}}^*.
\end{equation}
Thus $\Delta_{\mathrm{obs}}=\Delta_{\mathrm{within}}^*+\Delta_{\mathrm{comp}}^*$ exactly for the chosen standard. The observed contrast uses all eligible systems, whereas standardization uses shared system support. Consequently, $\Delta_{\mathrm{comp}}^*$ is a composition-associated residual that also includes the effect of this support restriction; it does not isolate reweighting on an identical support. This follows Kitagawa's rate-versus-composition logic~\cite{kitagawa1955components}, but is not the unique symmetric decomposition. Both components are descriptive and standard-dependent. A separately dated August registry check supplies historical retrieval context, not contemporary validation of September registries. Leave-one-system-out and joint exclusions test influence. These diagnostics do not redefine the estimand. Calculations use unrounded counts and rates; rounding occurs only in presentation.

\subsection{Failure Discipline, Provenance, and Validation}
\label{sec:provenance}
The resolver is fail-closed. Domain normalization never overwrites raw values; aliases retain evidence; parser failure terminates in a non-evaluable registry state; duplicate records reduce to one presence fact; and arbitrary fuzzy matching is excluded. Schema tests cover valid, malformed, confidential, duplicated, noncanonical, redirect, and alias cases. Aggregate invariants require every declaration to terminate in exactly one account or non-evaluable registry state and state totals to reconcile with population exports.

The pipeline emits artifact, declaration, registry, resolution, and population objects. Canonical population results can be traced from aggregate cells to normalized resolution rows, source hashes, line positions, parser versions, and collection runs. Separately dated sources have their own provenance: the June grouped fan-out export, the August registry-check log, historical removal inputs, and the surviving collection ledger are not silently treated as canonical resolution rows. Appendix~\ref{app:reproduction} specifies what each source can reproduce and its limits. {{Removal weights and review-surface annotations are attached only after transparency normalization. They support population assignment and bounded investigative illustrations, but cannot alter terminal resolution states or the RQ1 and RQ2 estimands.}}

\section{RQ1: Prevalence of the Seller-Identity Gap}\label{sec:prevalence}
{{RQ1 asks how much of the complete declared authorization surface fails the public seller-account check under the coverage-aware evidence rule, and how the resulting floor differs between reference and enforcement-linked populations.}}

\begin{table}[t]\centering\scriptsize\setlength{\tabcolsep}{4pt}
\caption{Coverage-robust seller-identity gap. Populations are cumulative; confidence intervals resample publisher hosts.}\label{tab:gradient}
\begin{tabular}{lrrr}\toprule
Pop. & Rows & Gap & 95\% CI\\\midrule
Ref. & 40,384,095 & 24.25\% & 23.97--24.51\%\\
$\geq100$ & 6,344,799 & 30.71\% & 29.98--31.40\%\\
$\geq1$k & 2,626,098 & 31.55\% & 29.87--32.97\%\\
$\geq10$k & 691,380 & 33.08\% & 31.40--34.56\%\\
\bottomrule
\end{tabular}\end{table}

\begin{figure}[t]\centering\includegraphics[width=0.6\figgradientwidth]{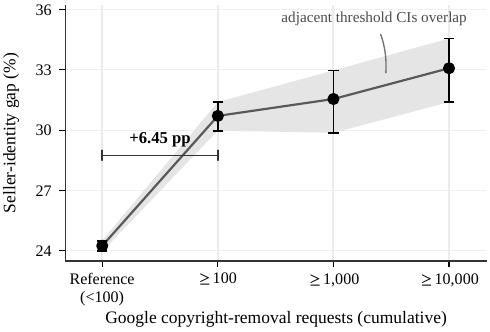}
\caption{Coverage-robust row-level floor. The primary contrast is the 6.45-point increase from the reference to $\geq100$; stricter-threshold intervals overlap.}\label{fig:gradient}\end{figure}

\subsection{Baseline and Enforcement-Linked Step}
Table~\ref{tab:gradient} and Figure~\ref{fig:gradient} show substantial non-resolution even in the reference: 24.25\% of 40,384,095 declaration rows. The floor rises to 30.71\%, 31.55\%, and 33.08\% in cumulative enforcement-linked populations. {{The reference estimate is central: almost one public authorization path in four stops before a public seller-account record, compared with almost one in three at the primary enforcement-linked threshold. The gap is therefore widespread rather than exclusive to enforcement-linked inventory.}}

The primary comparison is 6.45 percentage points, or 26.6\% relative to reference. Estimates then remain near one third. The reference interval separates from every enforcement-linked interval, whereas marginal intervals for adjacent enforcement-linked thresholds overlap. Overlap alone is not a test of a difference. A supplementary paired host bootstrap, sampling disjoint threshold-membership strata jointly (1,000 replicates, seed 1701), gives a $\geq1,000$ minus $\geq100$ contrast of 0.84 points, with a pointwise 95\% interval of $[-0.24,1.85]$. The evidence therefore supports a reference-to-enforcement-linked step, not a statistically separated increase at every threshold. {{Because the thresholds are nested observational strata, this pattern is not interpreted as a smooth dose-response relation or a causal effect; supplementary intervals are not adjusted for multiple comparisons.}}

The unit is a declaration row. Publishers with larger files contribute more weight, and one publisher may contain resolved and absent paths. A host-level any-gap indicator would erase the extent of mixed resolution. Conversely, row prevalence is not the share of publishers affected.

\subsection{Coverage, Uncertainty, and Selection}
Publisher bootstrap preserves each host's complete declaration set. Intervals widen as declaration-bearing hosts fall to 374 at the highest threshold. The ten largest hosts contribute 14.74\% of declaration weight in the $\geq1,000$ cohort, illustrating why host counts alone do not determine precision. Public-registry coverage ranges from 96.54\% to 95.08\%. This variation is smaller than the primary contrast. Unavailable registries cannot inflate a population's absent-only numerator, although differential availability can affect the contrast between floors. Coverage therefore remains a separate diagnostic rather than evidence of publisher opacity.

The comparison remains conditional on participation in transparency. Only 8.09\% of candidate reference hosts and 1.85--3.34\% of candidate enforcement-linked hosts expose usable declarations. The reference is low-removal rather than verified benign; thresholds are cumulative; and removal volume is an enforcement-attention signal. Percentages describe declaration rows among participating hosts, not all candidate hosts, monetized publishers, or publisher prevalence.

\subsection{Answer to RQ1}
{{The seller-identity gap is widespread across the observed declaration surface and higher in the selected enforcement-linked populations. It affects 24.25\% of reference declaration paths and 30.71\% at the primary enforcement-linked threshold, a 6.45-point step, then remains near one third at stricter thresholds. Under the stated evidence rule, these estimates quantify how often a visible authorization loses its minimum public account-verification path; they do not assign wrongdoing or responsibility.}}

\section{RQ2: Structure of the Enforcement-Linked Difference}\label{sec:structure}
{{RQ1 establishes a 6.45-point reference-to-$\geq100$ contrast. RQ2 asks where that difference lies across popularity support, within-system non-resolution, and ad-system composition. The analysis is descriptive: it localizes the contrast without interpreting the components as causes.}}

\subsection{Publisher Popularity}\label{sec:popularity}
Restricting enforcement-linked hosts to those in Tranco and comparing within common rank buckets leaves positive differences of +5.29, +7.48, and +8.92 points. These are differences from rank-matched references, not replacement gap estimates. Their persistence indicates that observed rank alone does not account for ordering on common support. {{The comparison is a sensitivity analysis over observed popularity. We exclude hosts outside Tranco, and unobserved differences in scale, maintenance, professionalization, or selection remain possible.}}

\subsection{Common-Standard Decomposition}\label{sec:decomposition}
\begin{figure}[t]\centering\includegraphics[width=0.6\columnwidth]{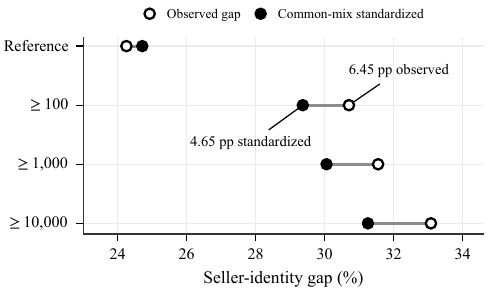}
\caption{Observed and common-mix standardized gaps. The primary contrast falls from 6.45 to 4.65 points under common system weights.}\label{fig:observed-standardized}\end{figure}

Standardized rates are 24.71\%, 29.37\%, 30.06\%, and 31.26\%. In the primary comparison, 4.65 points, approximately 72.1\%, remain as the within-system component. The remaining 1.80 points, approximately 27.9\%, form the ad-system-composition-associated residual. Together they reconstruct the 6.45-point observed contrast for the chosen standard.

{{Under common ad-system weights, most of the observed step remains within systems rather than disappearing with population composition. The remaining 1.80 points include both differences in declaration mix and the restriction to shared systems. On shared support alone, the observed contrast is 6.343 points: the support restriction contributes 0.109 points to the official residual, and reweighting on that support contributes 1.689 points. Shared systems cover 99.06\% of reference rows and 99.71\% of rows in the primary enforcement-linked cohort. We retain the official 1.80-point residual and refer to it below as the composition residual, with this support qualification. The allocation is standard-dependent and localizes the contrast; it does not identify mediation, mechanism, diligence, or responsibility~\cite{kitagawa1955components}.}}

\subsection{System Heterogeneity and Dated Registry Checks}\label{sec:system-heterogeneity}
\begin{figure}[t]\centering\includegraphics[width=0.6\figsystemswidth]{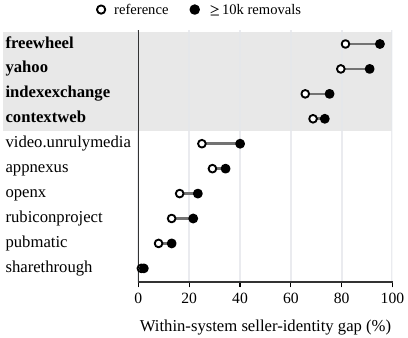}
\caption{Within-system gap for ten high-volume systems. Open points show the reference, and filled points show the cumulative $\geq10,000$ population.}\label{fig:systems}\end{figure}

Within-system floors range from near-complete resolution to approximately 66--100\% non-resolution in a smaller class of video, CTV, native, and retargeting systems. In many systems, the enforcement-linked endpoint exceeds reference, which explains why common-mix standardization does not eliminate the contrast. System values are descriptive and do not rank intent or responsibility. The ten systems in Figure~\ref{fig:systems} and the six excluded below form a fixed diagnostic panel, not a newly selected top-ten ranking.

\begin{table}[t]\centering\scriptsize
\caption{Distinct seller IDs in the canonical frozen registries, an independent August 13 count log, and canonical gap floors (\%). The August check is historical context, not a September validation.}\label{tab:livecheck}
\begin{tabular}{lrrrr}\toprule
System & Frozen & Aug. 13 & Ref. & $\geq10$k\\\midrule
contextweb & 46 & 46 & 68.7 & 73.4\\
yahoo & 377 & 381 & 79.7 & 91.0\\
criteo$^{*}$ & 1,652 & 1,651 & 92.3 & 99.6\\
outbrain & 2,915 & 1,000 & 84.8 & 90.3\\
freewheel & 2,260 & 2,250 & 81.5 & 95.1\\
indexexchange & 4,065 & 4,072 & 65.7 & 75.3\\\midrule
pubmatic & 6,515 & 6,476 & 7.9 & 13.1\\
sharethrough & 5,205 & 5,191 & 1.2 & 2.1\\\bottomrule
\end{tabular}
\parbox{0.95\columnwidth}{\scriptsize $^{*}$\texttt{criteo.com/sellers.json} redirected to \texttt{themediagrid.com}.}
\end{table}

An independently logged check on August 13, 2026 preserves eight distinct-ID counts, HTTP outcomes, and redirect destinations. Criteo's redirect is retained as an alias, not treated as absence. The September frozen lookup contains 2,915 Outbrain IDs; the historical count of 1,000 cannot validate its completeness. {{The dated log is a limited retrieval sanity check, not evidence of line-by-line equivalence or contemporary registry completeness. Original HTTP bodies were not retained by that probe, so a byte-for-byte historical replay is unavailable. Neither registry size nor the check explains why a particular coordinate is absent.}}

\subsection{Influence Diagnostics}
\begin{table*}[t]\centering\scriptsize
\caption{Robustness checks. Popularity entries are deltas from rank-matched references. Other entries are floors.}\label{tab:robustness}
\begin{tabular}{lrrrr}\toprule
Check & Ref. & $\geq100$ & $\geq1,000$ & $\geq10,000$\\\midrule
Observed & 24.25 & 30.71 & 31.55 & 33.08\\
Common mix & 24.71 & 29.37 & 30.06 & 31.26\\
Popularity delta & -- & +5.29 & +7.48 & +8.92\\\midrule
$-$freewheel & 22.86 & 28.82 & 29.69 & 31.13\\
$-$criteo & 23.86 & 30.26 & 31.08 & 32.59\\
$-$yahoo & 23.76 & 30.03 & 30.81 & 32.33\\
$-$indexexchange & 23.38 & 29.57 & 30.38 & 31.91\\
$-$outbrain & 24.07 & 30.53 & 31.37 & 32.87\\
$-$contextweb & 23.49 & 30.04 & 30.90 & 32.43\\
$-$all six & 19.88 & 25.33 & 26.06 & 27.40\\
\bottomrule
\end{tabular}\end{table*}

Dropping one panel system at a time preserves the step. Jointly dropping all six lowers absolute floors but leaves a 5.45-point primary contrast and the ordering of point estimates. {{These systems therefore influence the level of non-resolution, but neither a single system nor their joint exclusion accounts for the observed reference-to-enforcement-linked step. Point-estimate ordering is not evidence of statistical separation between adjacent thresholds.}}

\subsection{Answer to RQ2}
{{The enforcement-linked difference persists on common popularity support and under common ad-system weights. Of the 6.45-point primary contrast, 4.65 points, approximately 72\%, remain within systems, while 1.80 points, approximately 28\%, form the composition residual, including the small support-restriction term. Registry-sparse systems raise absolute floors but are not necessary for the population ordering. RQ2 therefore locates most of the difference within shared systems, with a smaller composition-associated remainder, without attributing either component to a causal mechanism or responsible actor.}}

\section{Ad Transparency Intelligence and Reviewable Evidence}
\label{sec:audit-surfaces}
{{RQ1 and RQ2 measure the seller-identity gap at population scale. We next illustrate evidence-bearing investigation without changing the estimand, common weights, or resolution states. We use \emph{Ad Transparency Intelligence} to describe this declaration-centered, ecosystem-level use of public advertising records: reconstructing and correlating declared supply-chain relationships to identify recurring coordinates, shared configurations, and failed public joins. The objective is not automatic attribution, but bounded public evidence that an analyst, intermediary, buyer, or regulator can inspect and corroborate. Fan-out comes from the separately archived June 21 study; shared-content membership and four case metrics use the canonical snapshot, with historical removal context identified separately. These illustrations are not additional prevalence samples.}}

\subsection{Recurring Seller-Account Coordinates}
\label{sec:account-concentration}
The June study groups declaration leads by $(\textit{artifact family},\textit{canonical ad system},\textit{seller ID})$. Keeping artifact family in the key avoids merging coordinates from different declaration mechanisms without evidence. We count archived grouped leads and distinct hosts; these leads are not interchangeable with raw normalized rows in the canonical snapshot. Identifier-based linkage is an established practice~\cite{starov2018dashboard,papadogiannakis2022publisherids,adalytics2021network}.
\begin{table}[t]
\centering
\small
\caption{Seller-account fan-out in the archived June 21 removal-linked deep-mapping study. It does not compare concentration with a matched reference or measure gap prevalence.}
\label{tab:fanout}
\begin{tabular}{lr}
\toprule
Metric & Value \\
\midrule
Seller-account clusters & 321,097 \\
Grouped declaration leads & 15,269,683 \\
Median hosts per account & 2 \\
99th percentile hosts & 902 \\
Maximum hosts & 12,703 \\
Fan-out Gini & 0.920 \\
Mixed DIRECT/RESELLER accounts & 30,947 \\
\bottomrule
\end{tabular}
\end{table}
{{The distribution compresses a large declaration surface into a smaller upper tail of reviewable coordinates. An ad system examining one high-fan-out coordinate can inspect a relationship declared across hundreds or thousands of removal-linked hosts; a buyer or investigator can prioritize coordinates affecting a large portion of the measured surface and identify the intermediary at which verification can continue. This is investigative leverage grounded in declared reach, not evidence of misconduct.}}
Fan-out includes resolved and absent coordinates. A seller account may legitimately serve many publishers, and resale can create broad reuse without concealing identity. Mixed relationship types can reflect legitimate differences in control, resale paths, or file maintenance. The maximum-fan-out account resolves correctly. This counterexample demonstrates that concentration and account absence are separate properties. The deep-mapping population has no matched concentration reference, so the Gini and upper tail cannot be attributed specifically to enforcement-linked publishers.

\subsection{Byte-Identical Declaration Configurations}
\label{sec:shared-configurations}
{{A complementary surface asks whether distinct publishers expose the same complete authorization configuration. We identify byte-identical declaration files by cryptographic hash. Unlike similarities inferred from names or external technical properties, byte identity preserves the exact declared coordinates and ordering.}}
\begin{table}[t]
\centering
\small
\caption{{{Shared-content case. The selective link is one byte-identical \texttt{ads.txt} configuration published by 36 domains.}}}
\label{tab:shared-content}
\begin{tabular}{lr}
\toprule
Property & Value \\
\midrule
Distinct domains & 36 \\
Shared content hashes & 1 \\
Accounts in shared file & 4,120 \\
Members with removal annotations & 97.2\% \\
Maximum removed URLs & 98,759 \\
\bottomrule
\end{tabular}
\end{table}
{{One fixed hash is observed on 36 domains publishing the same \texttt{ads.txt} with 4,120 accounts; 97.2\% of the matched historical member annotations carry removal volume. The maximum removed-URL count is historical context, not a new September enforcement observation. The canonical link is reproducible from the exact public bytes and exposes a common declaration configuration that can be inspected once and traced across all observed members. It identifies a focused review surface while retaining the underlying seller coordinates.}}
Exact byte identity establishes a shared declaration configuration, not common ownership, management, or coordinated misconduct. The same bytes may reflect common management, a shared publishing service, distributed templates, or copying. The defensible finding is configuration identity.

\subsection{Provenance-Bearing Case Packets}
\label{sec:investigation-cases}
The four fixed illustrative publishers have declaration and resolution metrics recomputed under the canonical absent-only floor. Their identities come from the historical case study, not from a search for extreme canonical outcomes. Historical copyright-removal weights remain context, not contemporary traffic or prevalence evidence. Canonical source hashes, declaration rows, registry provenance, alias decisions, and resolution states provide the available evidence path; the manifest distinguishes these from historical annotations and does not promise that canonical inputs alone can reconstruct every historical packet.
\begin{table*}[t]
\centering
\small
\caption{Four fixed anonymized cases with canonical declaration metrics and absent-only gap floors. Removed URLs retain historical context. The cases are not used to estimate population rates.}
\label{tab:cases}
\begin{tabular}{lrrrrrr}
\toprule
Case & Removed URLs & Declarations & Systems & Accounts & DIRECT rows & Gap \\
\midrule
High removal & 51.9M & 227 & 46 & 113 & 9 & 48.9\%\\
Broad surface 1 & 10.9M & 6,479 & 398 & 3,632 & 1,543 & 26.5\%\\
Broad surface 2 & 5.5M & 11,032 & 616 & 8,344 & 4,006 & 36.4\%\\
Sparse surface & 3.3M & 138 & 46 & 138 & 2 & 18.1\%\\
\bottomrule
\end{tabular}
\end{table*}
The examples include small declared surfaces with high removal volume and broad configurations that include both resolved and absent paths. Removal volume, declaration size, and gap fraction do not move mechanically in tandem. A host-level binary would lose this internal structure. The cases demonstrate traceability, not prevalence. Four selected publishers do not increase effective sample size or establish a case-level distribution.
{{Operationally, each packet identifies the exact publisher declaration, ad-system namespace, seller ID, registry state, and evidence provenance for further review. It turns an aggregate result into a bounded, reproducible investigative lead without extending inference beyond public records.}}

\subsection{What the Investigative Surfaces Add}
{{Recurring coordinates prioritize relationships by declared reach. Content identity identifies configurations replicated across publishers. Canonical case metrics connect selected hosts to row- and registry-level evidence. These structures offer complementary investigative paths, with distinct source dates and units; the historical fan-out export is not an automatic row-level drill-down from a canonical population cell. This declaration-native workflow supports correlation, corroboration, and review prioritization while preserving the distinction between a public lead and attribution.}}
None justifies automatic blocking, delisting, reputational labeling, ownership attribution, or a revenue claim. We do not evaluate analyst performance, workload, prioritization gain, or an end-to-end investigative workflow.

\section{Discussion and Implications}
\label{sec:discussion}
The results support one central interpretation: public participation in advertising transparency does not equal public accountability. A declared path becomes minimally accountable only when its namespace-safe coordinate resolves in the applicable available registry.

\subsection{Participation Is Not Accountability}
The important distinction is not visible versus invisible, but whether visible authorization terminates at the account record promised by the standards. This property is orthogonal to pooling: a high-fan-out account may resolve correctly, while a coordinate declared once may be absent. Account presence is necessary but insufficient for complete identity, legal-entity attribution, contract validity, execution, or payment.

The substantial 24.25\% reference floor means non-resolution is widespread rather than exclusive to enforcement-linked inventory. Conversely, higher floors in enforcement-linked populations do not imply that individual publishers or intermediaries caused the difference or acted improperly. {{The result identifies a supply-chain integrity problem: visible authorization metadata can remain syntactically usable while the reciprocal public account control fails. Buyers, exchanges, validators, brand-safety systems, regulators, and investigators can consume the declaration but cannot complete the verification path the public mechanism promises. This verifier-visible break is the security-measurement object studied here.}}

\subsection{State-Aware Conformance and Path-Specific Remediation}
A useful validator should preserve resolved, absent, confidential, aliased, malformed, unavailable, and otherwise non-evaluable outcomes rather than collapse them into pass/fail. An absent ID under an available registry supports account review; an unavailable registry supports endpoint remediation and retry; a confidential record preserves account presence without disclosure; an alias should require a versioned rule.

\begin{table}[t]
\centering
\small
\setlength{\tabcolsep}{4pt}
\caption{Bounded follow-up actions at the public seller-account boundary.}
\label{tab:actions}
\begin{tabular}{@{}p{0.23\columnwidth}p{0.72\columnwidth}@{}}
\toprule
Actor & Bounded action \\
\midrule
Ad system & Reconcile declared coordinates; publish the account or an explicit invalid/stale state \\
Publisher & Remove stale rows and obtain current coordinates \\
Buyer / DSP & Prioritize unresolved paths without equating resale or concentration with abuse \\
Standards body & Define outcomes for presence, confidentiality, aliases, malformed registries, and unavailable evidence \\
Researcher / regulator & Measure recurrence and remediation under comparable populations, versions, coverage, and provenance \\
\bottomrule
\end{tabular}
\end{table}

{{The check can run without inspecting personal data or creative content and can act on an affected seller path rather than the entire publisher. Table~\ref{tab:actions} maps each evidence state to a bounded follow-up action: reconciliation for a missing account, endpoint repair for unavailable evidence, disclosure-aware handling for confidential records, and versioned treatment for aliases. These are implementable implications of the state model, not evaluated policies. We do not estimate implementation cost, false-positive burden, analyst performance, adoption feasibility, or auction effects.}}

\subsection{Interpreting the Decomposition}
The 4.65-point within-system component points to follow-up on account and registry reconciliation within shared systems. The 1.80-point composition-associated residual directs attention to declaration mix and the small contribution of the shared-support restriction. Neither identifies mechanism or responsibility. Aggregate change can reflect resolved accounts, removed declarations, new aliases, composition change, reduced participation, or lost coverage. A lower floor caused only by lost registry coverage is not an improvement.

{{The distinction also provides a more informative basis for future longitudinal evaluation. Improvement within an ad system should primarily reduce the within-system component, whereas changes in the intermediaries used by publishers or buyers should primarily affect the composition residual. Tracking both quantities can distinguish repair of public account resolution from a shift in declaration mix, provided populations, common support, common weights, parser rules, and registry coverage remain comparable, and support changes are reported separately.}}

\subsection{Ad Transparency Intelligence Without Automatic Attribution}
{{Recurring coordinates, byte-identical files, and provenance-bearing case packets turn a large integrity surface into reviewable evidence. In the Ad Transparency Intelligence framing introduced above, these surfaces support relationship reconstruction, correlation, corroboration, and review prioritization by showing which seller coordinates recur, which complete configurations are shared, where public joins fail, and which intermediary provides the next verifiable point of inquiry.}}

{{This investigative value does not require equating correlation with attribution. Fan-out includes resolved and absent coordinates; the maximum-fan-out account resolves. Byte-identical files establish configuration identity but may result from shared services, templates, or copying. Case packets preserve the exact evidence path but do not reveal intent or beneficial ownership. The surfaces therefore support targeted corroboration rather than automatic blocking, delisting, reputational labeling, or claims of common control~\cite{starov2018dashboard,papadogiannakis2022publisherids,adalytics2021network}.}}

\subsection{Remediation and the Declared--Observed Research Boundary}
Prior notification work supplies a natural intervention-study design~\cite{vekaria2024notifications}. Vekaria et al. used contact information exposed through \texttt{ads.txt} and \texttt{sellers.json} to reach relevant stakeholders, providing a concrete precedent for a future notification study. Our paper neither sends notifications nor evaluates their effectiveness, and we do not assume public contact information exists, is current, or is sufficient for remediation. A future snapshot can distinguish newly published accounts, publisher-removed declarations, documented aliases, recovered availability, and persistent gaps. Comparable populations, parser versions, canonicalization rules, alias rules, evidence states, common weights, and registry coverage are necessary so lost observability isn't mistaken for improvement.

{{A second research direction is systematic comparison between declared relationships and live RTB evidence. Reconstructing the public seller-account graph establishes which paths are declared and publicly resolvable; the next question is which of those paths become observable in request routing, bidding, winning, billable impressions, settlement, or payout. These stages require a dedicated sampling frame, coverage model, and phase-aware identifiers, and must remain analytically distinct~\cite{pastor2021adschain,zhong2023addax,mrc2026auction}. The present paper establishes the declaration-side prerequisite for that comparison rather than anticipating its empirical results.}}

\section{Limitations}\label{sec:limitations}
{{The study measures a specific public seller-account property under explicit construct, population, temporal, statistical, and review boundaries. These limits define the inference supported by the declaration-side evidence rather than reducing the value of the public-account check.}}

\subsection{Construct and Attribution Boundaries}
The construct ends at exact public account presence. A present ID does not validate name, domain, legal entity, beneficial owner, contract, authorization at auction time, or activity. A confidential record establishes presence while withholding identity. An absent ID does not prove that no private account or contract exists. Public artifacts reveal a state but not whether it stems from a stale or copied declaration, migration, error, registry omission, unpublished private account, or fabrication. Pooling, relationship mismatch, account absence, identity disclosure, execution, and payment are distinct. {{Fan-out, mixed relationships, byte-identical files, and case packets are declaration-native review signals. They can identify recurring coordinates, shared configurations, and exact evidence paths, but do not establish ownership, coordination, intent, or abuse.}}

\subsection{Population, Participation, and Temporal Scope}
The reference is high-popularity and low-removal, not verified benign. Copyright-removal counts reflect enforcement attention and can depend on scale, visibility, rights-holder effort, reporting practice, and duplicated URLs. Populations are cumulative observational strata, not treatments. Only hosts exposing usable declarations contribute rows. Participation differs substantially across candidate populations. Results characterize participating hosts' declared surface, not all candidates, all removal targets, all monetized publishers, or the share of publishers affected. Row weighting gives more weight to declaration-heavy publishers. Host-level, unique-coordinate, \texttt{DIRECT}-only, and conditional estimands answer different questions.

The estimator uses a frozen snapshot assembled from observations with different timestamps, not a simultaneous crawl. Declarations, registries, redirects, aliases, confidentiality, and states change over time. The August count log and June fan-out study are separately dated and do not validate September completeness or establish ecosystem-wide stability, persistence, incidence, or remediation. The count-only August probe cannot be replayed byte for byte because it did not retain response bodies.

\subsection{Measurement and Statistical Uncertainty}
Unavailable registries remain in the denominator and cannot enter the numerator, making that source of missingness conservative. Other errors need not be one-sided: alias, canonicalization, parser, or materialization errors can move rows between states. Raw values, hashes, bytes, versions, tests, and reason codes mitigate but do not eliminate these risks.

Publisher-clustered bootstrap intervals address within-host dependence for the row estimand~\cite{cameron2008bootstrap}. Publishers may share files, accounts, providers, templates, or management, so not all ecosystem dependence is captured. Popularity matching addresses observed Tranco support only. {{The common-standard decomposition is descriptive and depends on the selected weights and shared support. The 4.65-point within-system component and 1.80-point composition residual sum to the 6.45-point observed contrast under that standard; the residual includes 0.109 points from support restriction. These components do not estimate mediation or causality~\cite{kitagawa1955components}.}}

\subsection{Review-Surface and Case Boundaries}
The historical deep-mapping population has no matched concentration reference. Its fan-out Gini and upper tail do not show that concentration is distinctive of enforcement-linked inventory or predictive of absence. The maximum-fan-out account was resolved in that dated study. Four cases illustrate available canonical evidence paths, not prevalence, ground truth, or increased sample size. Historical removal context is not a new measurement of enforcement activity. Analyst accuracy, workload, cost, prioritization gain, false-positive burden, notification, and remediation are not evaluated.

\subsection{Boundary with Observable Execution}
{{Public declarations record claimed authorization relationships; they do not establish that a seller account participated in a live auction or received economic benefit. Request routing, returned bids, wins, billable impressions, settlement, and publisher payout are separate evidentiary stages. A future declared--observed study must model their sampling, coverage, identifiers, and inference independently rather than treating browser-visible traffic as a direct validation of the declaration-row floor.}}

\paragraph{Scope of claims.}
{{On the frozen transparency-participating surface, a substantial fraction of authorizations point to seller IDs absent from applicable available registries, and the floor is higher in the selected enforcement-linked populations. The paper quantifies and localizes this public-account failure and connects it to reviewable declaration evidence. It does not establish why an account is absent, who is responsible, whether review signals indicate common control, whether a declared path executed in live advertising traffic, or whether money changed hands.}}

\section{Conclusion}\label{sec:conclusion}
{{Public advertising transparency becomes accountable only when a seller coordinate declared by a publisher resolves to the account record promised by the applicable available registry. We formalize failure of this minimum property as the seller-identity gap and measure it using 233,416,998 normalized advertising-transparency records under an absent-only, coverage-aware evidence rule.}}

{{The result is substantial: the gap affects 24.25\% of declaration paths in the high-popularity, low-removal reference and 30.71\% at the primary enforcement-linked threshold, a 6.45-point step, before remaining near one third at stricter thresholds. Under common weights over 1,469 ad systems, 4.65 points, approximately 72\%, remain within systems, while 1.80 points, approximately 28\%, form the composition residual, including the small support-restriction term. The decomposition locates most of the observed difference within shared systems, with a smaller composition-associated remainder, without assigning cause or responsibility.}}

{{Recurring seller-account coordinates, byte-identical declaration configurations, and provenance-bearing case packets translate these population results into reviewable evidence. We frame this declaration-centered investigative use of public advertising records as Ad Transparency Intelligence: a reproducible way to reconstruct declared relationships, identify where public verification stops, and prioritize corroboration. These surfaces support investigation without converting correlation into ownership, intent, misconduct, or economic attribution.}}

{{The resulting contribution is a controlled path from a public security property, to population-scale measurement, to comparative localization, and finally to the exact declaration and registry evidence that produced the result. The next step is a systematic comparison with live RTB traffic, which requires its own sampling frame, coverage model, and phase-aware identifiers. Advertising transparency supports accountability and investigation only when declarations remain publicly verifiable; measuring where that verification fails turns a nominal transparency mechanism into a reviewable security surface.}}

\appendix
\section{Ethical Considerations}\label{app:ethics}
{{This study analyzes public advertising-transparency artifacts to measure public seller-account non-resolution and construct declaration-native review surfaces. The design minimizes operational impact and separates reproducible findings from claims about identity, intent, ownership, coordination, or economic outcome. An absent public record is a verifier-visible state, not a characterization of an actor.}}

\paragraph{Public-data acquisition and load minimization.}
Collection is rate-limited and preserves retrieval status, timestamps, redirects, hashes, source runs, and parser provenance. Repeated requests are not substitutes for missing evidence. The study does not bypass authentication, purchase advertising, click ads, submit conversions, or interfere with auctions. Failed or challenged retrievals remain explicit measurement states rather than triggering aggressive recovery.

\paragraph{Data minimization and sensitive fields.}
{{The pipeline retains public declaration and registry artifacts, technical metadata, hashes, account coordinates, and provenance needed for resolution and verification. The pipeline processes exact identifiers and source records only as needed to reproduce the public-account check, construct review surfaces, and support controlled case validation. Materials that are unnecessary for verification are aggregated, transformed, restricted, or withheld. Public availability does not remove the obligation to minimize redistribution risk.}}

\paragraph{Population labels and reporting.}
Removal information constructs observational populations and is not ground truth for infringement, intent, seller behavior, or intermediary knowledge. Results are reported in aggregate.\color{black}{Cases are reported in anonymized form, with exact identifiers retained in a supervised evidence key.} The fixed historical case identities are independent of the canonical population results; their metrics are recomputed without outcome-based replacement. Neither the selected cases nor the deep-mapping review surfaces constitute additional prevalence observations.

\paragraph{Responsible disclosure.}
Any notification is scoped to verifiable path, namespace, seller ID, registry state, source hash, line position, retrieval provenance, and reason code. It must distinguish absent from unavailable, malformed, confidential, aliased, and non-evaluable states. It does not claim fraud, infringement, common ownership, contractual non-authorization, revenue, or payment without independent evidence. The manuscript must describe only disclosure actually completed before submission.

\paragraph{Risk of misuse.}
{{Recurring coordinates, byte-identical declaration configurations, and case packets can support legitimate review but may be misinterpreted as blocklists or attribution evidence. We mitigate this risk by preserving terminal states, presenting alternative explanations, anonymizing cases during review, and recommending path-specific corroboration rather than automatic blocking, delisting, or reputational labeling. The artifact release applies the same boundary: enough provenance to reproduce the aggregate findings and verify selected cases, without turning investigative leads into unsupported accusations.}}

\section{Open Science}
\label{app:open-science}
{{The accompanying artifact supports verification of the public-data measurements, common-standard decomposition, and declaration-native investigative surfaces while respecting the disclosure boundaries in Appendix~\ref{app:ethics}. Availability and reproducibility claims in this paper are limited to the materials identified in the artifact manifest.}}

\subsection{Artifact Contents and Traceability}
\label{app:artifact-contents}
The release distinguishes five linked object families:
\begin{itemize}
\item \textbf{Artifact objects}: requested URLs, redirect chains, outcomes, timestamps, hashes, artifact families, parser versions, and source runs;
\item \textbf{Declaration objects}: hosts, source lines, raw and canonical ad-system domains, seller IDs, relationship types, and source provenance;
\item \textbf{Registry objects}: seller records, availability and parser states, confidentiality indicators, hashes, and provenance;
\item \textbf{Resolution objects}: declaration-to-registry links, alias decisions, terminal states, and reason codes;
\item \textbf{Population objects}: frozen host labels and the corresponding removal and popularity inputs.
\end{itemize}
The artifact manifest identifies schemas, configurations, derived objects, analysis code, versions, file sizes, and cryptographic digests, and assigns each aggregate to its actual source. Canonical population aggregates are traceable to normalized resolution rows and, where disclosure permits, source hashes, line positions, parser versions, alias rules, and collection runs. The historical grouped fan-out export and count-only registry log have different reproducibility boundaries. Complete canonical exports, rather than selected case packets, remain the source of population estimates.

\subsection{Reproducing the Main Claims}
\label{app:reproduction}
{{Canonical inputs support Tables~\ref{tab:populations}, \ref{tab:gradient}, \ref{tab:robustness}, \ref{tab:systems}, and \ref{tab:systems-cont}: RQ1 denominators, absent-only numerators, registry coverage and host-bootstrap intervals, and RQ2 within-system floors, common weights, standardized rates, and the 4.65/1.80-point decomposition. Tables~\ref{tab:states} and \ref{tab:actions} describe the method and bounded follow-up actions rather than empirical aggregates.}}

Separately dated sources are not attributed to the canonical snapshot. Table~\ref{tab:livecheck} combines canonical frozen-registry counts and floors with the independently logged August 13 retrieval counts; the log reproduces eight counts and retrieval outcomes, but not original response bodies. Table~\ref{tab:fanout} uses the archived June 21 grouped lead export: its cluster distribution and Gini can be recomputed from that export, not from a claim that every grouped lead is a canonical resolution row. Table~\ref{tab:shared-content} revalidates a fixed exact-content group against canonical observations, retaining historical removal context. Table~\ref{tab:cases} recomputes the declaration and resolution metrics for the four fixed case hosts; historical removal weights come from a separate source. Table~\ref{tab:collection-scale} sums the surviving lane summaries deduplicated by lane and run identifier at the declared cutoff. It does not establish that every past acquisition is retained.

The provenance manifest preserves source dates, hashes, units, derivation rules, and access conditions. The June full-frontier export lacks a historical output digest; a digest computed during recovery identifies the preserved file, not contemporaneous proof of its completeness. Workflows keep acquisition quantities, normalized objects, grouped leads, declaration rows, resolution states, population estimates, and review objects distinct, and report unavailable evidence explicitly.

\subsection{Review-Surface and Controlled Materials}
\label{app:sensitive-materials}
{{The releasable review materials comprise schemas, derivation code, aggregate fan-out statistics, exact-content grouping logic, anonymized case structures, and non-sensitive provenance fields identified in the artifact manifest. We transform, restrict, or withhold exact operational identifiers, anonymization keys, supervised case mappings, and materials whose release could enable unsupported attribution. The manifest records the treatment and disclosure rationale applied to each non-public material.}}

\FloatBarrier
\Needspace{0.35\textheight}
\subsection{Supporting Measurement Results}
\label{app:additional-results}
\begin{table}[H]
\centering
\scriptsize
\setlength{\tabcolsep}{3.5pt}
\caption{Within-system floors for five fixed high-volume diagnostic systems. The Rows column sums population memberships, including nested cohorts, not globally distinct declarations. Values do not rank responsibility or culpability.}
\label{tab:systems}
\begin{tabular}{lrrrrr}
\toprule
System & Rows & Ref. & $\geq100$ & $\geq1$k & $\geq10$k \\
\midrule
pubmatic & 4,084,298 & 7.94\% & 11.95\% & 11.94\% & 13.10\%\\
rubiconproject & 3,205,965 & 13.08\% & 18.24\% & 18.87\% & 21.58\%\\
appnexus & 2,480,941 & 29.15\% & 32.83\% & 32.78\% & 34.32\%\\
openx & 1,841,776 & 16.25\% & 20.46\% & 20.74\% & 23.40\%\\
video.unrulymedia & 1,413,551 & 25.00\% & 36.31\% & 37.83\% & 40.04\%\\
\bottomrule
\end{tabular}
\end{table}

\begin{table}[H]
\centering
\scriptsize
\setlength{\tabcolsep}{3.5pt}
\caption{Within-system floors for the remaining five high-volume systems. This table continues Table~\ref{tab:systems}.}
\label{tab:systems-cont}
\begin{tabular}{lrrrrr}
\toprule
System & Rows & Ref. & $\geq100$ & $\geq1$k & $\geq10$k \\
\midrule
freewheel & 1,247,843 & 81.52\% & 92.23\% & 93.37\% & 95.08\%\\
indexexchange & 1,084,825 & 65.68\% & 73.33\% & 73.56\% & 75.26\%\\
sharethrough & 1,050,204 & 1.17\% & 1.64\% & 1.64\% & 2.13\%\\
contextweb & 846,964 & 68.75\% & 69.78\% & 70.46\% & 73.37\%\\
yahoo & 473,413 & 79.68\% & 87.38\% & 88.02\% & 91.04\%\\
\bottomrule
\end{tabular}
\end{table}

\begin{table}[H]
\centering
\scriptsize
\setlength{\tabcolsep}{3.2pt}
\caption{Surviving transparency campaign summaries, April 16--September 21, deduplicated by lane and run ID at September 28, 08:30:43 UTC. Acquisition quantities do not enter gap denominators or prove a complete acquisition census.}
\label{tab:collection-scale}
\begin{tabular}{lrrrr}
\toprule
Artifact & Runs & Fetches & Candidates & Valid \\
\midrule
\texttt{ads.txt} & 1,422 & 34,389,408 & 28,208,567 & 18,252,554\\
\texttt{app-ads.txt} & 1,179 & 13,611,691 & 11,020,427 & 5,403,969\\
\texttt{sellers.json} & 706 & 3,220,992 & 2,647,009 & 343,655\\
\midrule
Total & 3,307 & 51,222,091 & 41,876,003 & 24,000,178\\
\bottomrule
\end{tabular}
\end{table}

\FloatBarrier
\subsection{Release Statement}
\label{app:release-statement}
The artifact manifest records the repository or archive, version, license, and access conditions used for the submission. It specifies which materials are released, transformed, restricted, or withheld and provides the corresponding disclosure rationale. Claims of artifact availability, functionality, and reproducibility in this paper apply only to the contents and access conditions recorded in that manifest.

\section*{Author Note}
The affiliation identifies Tommaso Grotto's current professional role. His contribution to this research was undertaken in a personal capacity.

\bibliographystyle{unsrtnat}
\bibliography{references}

@misc{adalytics2021network,
  author = {{Adalytics}},
  title = {{Using Network Analysis to Identify Related Websites}},
  year = {2021}
}

@misc{ahmad2024misinformation,
  author = {Ahmad, Wajeeha and Sen, Ananya and Eesley, Charles and Brynjolfsson, Erik},
  title = {{Companies Inadvertently Fund Online Misinformation Despite Consumer Backlash}},
  year = {2024}
}

@misc{alrwais2017bulletproof,
  author = {Alrwais, Sumayah and others},
  title = {{Under the Shadow of Sunshine: Understanding and Detecting Bulletproof Hosting on Legitimate Service Provider Networks}},
  year = {2017}
}

@misc{ana2023mfa,
  author = {{Association of National Advertisers}},
  title = {{Programmatic Media Supply Chain Transparency Study}},
  year = {2023}
}

@misc{andreou2018adtransparency,
  author = {Andreou, Athanasios and others},
  title = {{Investigating Ad Transparency Mechanisms in Social Media: A Case Study of Facebook Explanations}},
  year = {2018}
}

@misc{angus2024adobservatory,
  author = {Angus, Daniel and others},
  title = {{The Australian Ad Observatory: Technical and Data Report}},
  year = {2024}
}

@misc{angus2024observability,
  author = {Angus, Daniel and others},
  title = {{Computational Methods for Improving the Observability of Platform-Based Advertising}},
  year = {2024}
}

@misc{aqeel2020headerbidding,
  author = {Aqeel, Waqar and others},
  title = {{Untangling Header Bidding Lore: Some Myths, Some Truths, and Some Hope}},
  year = {2020}
}

@misc{bashir2016retargeting,
  author = {Bashir, Muhammad Ahmad and Arshad, Sajjad and Robertson, William and Wilson, Christo},
  title = {{Tracing Information Flows Between Ad Exchanges Using Retargeted Ads}},
  year = {2016}
}

@misc{bashir2018diffusion,
  author = {Bashir, Muhammad Ahmad and Wilson, Christo},
  title = {{Diffusion of User Tracking Data in the Online Advertising Ecosystem}},
  year = {2018}
}

@misc{bashir2019adstxt,
  author = {Bashir, Muhammad Ahmad and Arshad, Sajjad and Kirda, Engin and Robertson, William and Wilson, Christo},
  title = {{A Longitudinal Analysis of the ads.txt Standard}},
  year = {2019}
}

@misc{batikas2019followmoney,
  author = {Batikas, Michail and Claussen, Jorg and Peukert, Christian},
  title = {{Follow the Money: Online Piracy and Self-Regulation in the Advertising Industry}},
  year = {2019}
}

@misc{benzaamia2026dsa,
  author = {Benzaamia, Abir and El Fraihi, Asmaa and Abdelaziz, Ines and Goga, Oana},
  title = {{A Year Under the DSA: Ad Transparency Uneven Landscape}},
  year = {2026}
}

@misc{bozarth2021marketforces,
  author = {Bozarth, Lia and Budak, Ceren},
  title = {{Market Forces: Quantifying the Role of Top Credible Ad Servers in the Fake News Ecosystem}},
  year = {2021}
}

@misc{cameron2008bootstrap,
  author = {Cameron, A. Colin and Gelbach, Jonah B. and Miller, Douglas L.},
  title = {{Bootstrap-Based Improvements for Inference with Clustered Errors}},
  year = {2008}
}

@misc{cangialosi2016keysharing,
  author = {Cangialosi, Frank and others},
  title = {{Measurement and Analysis of Private Key Sharing in the HTTPS Ecosystem}},
  year = {2016}
}

@misc{chachra2015affiliate,
  author = {Chachra, Neha and Savage, Stefan and Voelker, Geoffrey M.},
  title = {{Affiliate Crookies: Characterizing Affiliate Marketing Abuse}},
  year = {2015}
}

@misc{cook2020kashf,
  author = {Cook, John and Nithyanand, Rishab and Shafiq, Zubair},
  title = {{Inferring Tracker-Advertiser Relationships in the Online Advertising Ecosystem Using Header Bidding}},
  year = {2020}
}

@misc{dave2012clickspam,
  author = {Dave, Vacha and Guha, Saikat and Zhang, Yin},
  title = {{Measuring and Fingerprinting Click-Spam in Ad Networks}},
  year = {2012}
}

@misc{dca2021breakingbads,
  author = {Digital Citizens Alliance and White Bullet},
  title = {{Breaking (B)ads: How Advertiser-Supported Piracy Helps Fuel a Booming Illegal Market}},
  year = {2021}
}

@misc{dca2022ransomware,
  author = {Digital Citizens Alliance and White Bullet},
  title = {{From Piracy to Ads to Ransomware: How Illicit Actors Use Digital Ads on Piracy Sites}},
  year = {2022}
}

@misc{englehardt2016openwpm,
  author = {Englehardt, Steven and Narayanan, Arvind},
  title = {{Online Tracking: A 1-Million-Site Measurement and Analysis}},
  year = {2016}
}

@misc{euipo2026iprads,
  author = {European Union Intellectual Property Office and White Bullet},
  title = {{Online Advertising on IPR-Infringing Websites and Apps 2025}},
  year = {2026}
}

@misc{google_transparency,
  author = {{Google}},
  title = {{Content Delistings Due to Copyright: Google Transparency Report}},
  year = {2026}
}

@misc{googlewhiteops2018_3ve,
  author = {Google and White Ops},
  title = {{The Hunt for 3ve: Taking Down a Major Ad Fraud Operation}},
  year = {2018}
}

@misc{han2022infrastructure,
  author = {Han, Catherine and Kumar, Deepak and Durumeric, Zakir},
  title = {{On the Infrastructure Providers That Support Misinformation Websites}},
  year = {2022}
}

@misc{hounsel2020infrastructure,
  author = {Hounsel, Austin and others},
  title = {{Identifying Disinformation Websites Using Infrastructure Features}},
  year = {2020}
}

@misc{iab_adcom,
  author = {{IAB Tech Lab}},
  title = {{AdCOM: Advertising Common Object Model}},
  year = {2024}
}

@misc{iab_adstxt,
  author = {{IAB Tech Lab}},
  title = {{Authorized Digital Sellers (ads.txt), Specification v1.1}},
  year = {2022}
}

@misc{iab_appadstxt,
  author = {{IAB Tech Lab}},
  title = {{app-ads.txt: Authorized Sellers for Apps}},
  year = {2019}
}

@misc{iab_openrtb,
  author = {{IAB Tech Lab}},
  title = {{OpenRTB Specification v2.6}},
  year = {2022}
}

@misc{iab_sellersjson,
  author = {{IAB Tech Lab}},
  title = {{sellers.json and the OpenRTB SupplyChain Object}},
  year = {2019}
}

@misc{iab_transparency_center,
  author = {{IAB Tech Lab}},
  title = {{Transparency Center and Supply-Chain Validation Tools}},
  year = {2026}
}

@misc{iab_vast43,
  author = {{IAB Tech Lab}},
  title = {{Digital Video Ad Serving Template (VAST) v4.3}},
  year = {2022}
}

@misc{iabeurope2025adoption,
  author = {{IAB Europe}},
  title = {{Adoption of Supply Chain Transparency Standards in Europe}},
  year = {2025}
}

@misc{iabey2015untrustworthy,
  author = {Interactive Advertising Bureau and Ernst and Young},
  title = {{What Is an Untrustworthy Supply Chain Costing the U.S. Digital Advertising Industry?}},
  year = {2015}
}

@misc{isba2020supplychain,
  author = {ISBA and Association of Online Publishers and PwC},
  title = {{Programmatic Supply Chain Transparency Study}},
  year = {2020}
}

@misc{ivp5adtechaudit,
  author = {{ivp5}},
  title = {{ads.txt Framework Integrity Audit}},
  year = {2026}
}

@misc{jounce2025reselling,
  author = {{Jounce Media}},
  title = {{The Marketer Guide to Programmatic Reselling}},
  year = {2025}
}

@misc{kitagawa1955components,
  author = {Kitagawa, Evelyn M.},
  title = {{Components of a Difference Between Two Rates}},
  year = {1955}
}

@misc{kline2018placement,
  author = {Kline, Jeffery and Cahn, Aaron and Barford, Paul},
  title = {{Placement Laundering and the Complexities of Attribution in Online Advertising}},
  year = {2018}
}

@misc{korolova2011privacy,
  author = {Korolova, Aleksandra},
  title = {{Privacy Violations Using Microtargeted Ads: A Case Study}},
  year = {2011}
}

@misc{kroll2023qualitative,
  author = {{Kroll}},
  title = {{ANA Programmatic Media Supply Chain Transparency Study: Qualitative Insights}},
  year = {2023}
}

@misc{lepochat2019tranco,
  author = {Le Pochat, Victor and others},
  title = {{Tranco: A Research-Oriented Top Sites Ranking Hardened Against Manipulation}},
  year = {2019}
}

@misc{li2012madtracer,
  author = {Li, Zhou and others},
  title = {{Knowing Your Enemy: Understanding and Detecting Malicious Web Advertising}},
  year = {2012}
}

@misc{lu2021whowas,
  author = {Lu, Chaoyi and others},
  title = {{From WHOIS to WHOWAS: A Large-Scale Measurement Study of Domain Registration Privacy under the GDPR}},
  year = {2021}
}

@misc{moss2021darkpool,
  author = {Moss, Rocky},
  title = {{Further Investigation into the Dark Pool Sales House Phenomenon}},
  year = {2021}
}

@misc{moss2021deepsee,
  author = {Moss, Rocky},
  title = {{Evaluating the Ecosystem: What We Have Learned by Matching ads.txt Entries to sellers.json Files}},
  year = {2021}
}

@misc{mozilla2024fulldisclosure,
  author = {Mozilla Foundation and CheckFirst},
  title = {{Full Disclosure: Stress Testing Tech Platforms Ad Repositories}},
  year = {2024}
}

@misc{mrc2026auction,
  author = {{Media Rating Council}},
  title = {{Digital Advertising Auction Transparency Standards}},
  year = {2026}
}

@misc{musa2022atom,
  author = {Musa, Maaz Bin and Nithyanand, Rishab},
  title = {{ATOM: Ad-Network Tomography}},
  year = {2022}
}

@misc{oh2022cartology,
  author = {Oh, Cumhur and Kanich, Chris and McCoy, Damon and Pearce, Paul},
  title = {{Cart-ology: Intercepting Targeted Advertising via Ad Network Identity Entanglement}},
  year = {2022}
}

@misc{olejnik2014auction,
  author = {Olejnik, Lukasz and Tran, Minh-Dung and Castelluccia, Claude},
  title = {{Selling Off Privacy at Auction}},
  year = {2014}
}

@misc{pachilakis2019waterfalls,
  author = {Pachilakis, Michalis and Papadopoulos, Panagiotis and Markatos, Evangelos P. and Kourtellis, Nicolas},
  title = {{No More Chasing Waterfalls: A Measurement Study of the Header Bidding Ad-Ecosystem}},
  year = {2019}
}

@misc{papadogiannakis2021postcookie,
  author = {Papadogiannakis, Emmanouil and others},
  title = {{User Tracking in the Post-Cookie Era: How Websites Bypass GDPR Consent to Track Users}},
  year = {2021}
}

@misc{papadogiannakis2022publisherids,
  author = {Papadogiannakis, Emmanouil and others},
  title = {{Leveraging Google Publisher-Specific IDs to Detect Website Administration}},
  year = {2022}
}

@misc{papadogiannakis2023whofunds,
  author = {Papadogiannakis, Emmanouil and others},
  title = {{Who Funds Misinformation? A Systematic Analysis of the Ad-Related Profit Routines of Fake News Sites}},
  year = {2023}
}

@misc{papadogiannakis2024adlaundering,
  author = {Papadogiannakis, Emmanouil and others},
  title = {{Ad Laundering: How Websites Deceive Advertisers into Rendering Ads Next to Illicit Content}},
  year = {2024}
}

@misc{papadogiannakis2025beforeafter,
  author = {Papadogiannakis, Emmanouil and others},
  title = {{Before and After: The Effect of EU 2022 Code of Practice on Disinformation}},
  year = {2025}
}

@misc{papadogiannakis2025darkside,
  author = {Papadogiannakis, Emmanouil and Kourtellis, Nicolas and Papadopoulos, Panagiotis and Markatos, Evangelos P.},
  title = {{Welcome to the Dark Side: Analyzing the Revenue Flows of Fraud in the Online Ad Ecosystem}},
  year = {2025}
}

@misc{papadopoulos2017product,
  author = {Papadopoulos, Panagiotis and others},
  title = {{If You Are Not Paying for It, You Are the Product: How Much Do Advertisers Pay to Reach You?}},
  year = {2017}
}

@misc{pastor2021adschain,
  author = {Pastor, Antonio and Cuevas, Ruben and Cuevas, Angel and Azcorra, Arturo},
  title = {{Establishing Trust in Online Advertising with Signed Transactions}},
  year = {2021}
}

@misc{pearce2014zeroaccess,
  author = {Pearce, Paul and others},
  title = {{Characterizing Large-Scale Click Fraud in ZeroAccess}},
  year = {2014}
}

@misc{poilve2020adweb,
  author = {Poilve, Benjamin},
  title = {{Discover the Advertising Web with Sellers.json Files}},
  year = {2020}
}

@misc{rafique2016freestreaming,
  author = {Rafique, M. Zubair and others},
  title = {{It Is Free for a Reason: Exploring the Ecosystem of Free Live Streaming Services}},
  year = {2016}
}

@misc{sanchezrola2021cookies,
  author = {Sanchez-Rola, Iskander and others},
  title = {{Journey to the Center of the Cookie Ecosystem: Unraveling Actors Roles and Relationships}},
  year = {2021}
}

@misc{sheaib2025unmasking,
  author = {Sheaib, Hussein and Feldmann, Anja and Dao, Ha},
  title = {{Unmasking the Shadows: A Cross-Country Study of Online Tracking in Illegal Movie Streaming Services}},
  year = {2025}
}

@misc{siby2022webgraph,
  author = {Siby, Sandra and others},
  title = {{WebGraph: Capturing Advertising and Tracking Information Flows for Robust Blocking}},
  year = {2022}
}

@misc{smith2020correlation,
  author = {Smith, Kyle},
  title = {{Correlation Study Between ads.txt and sellers.json}},
  year = {2020}
}

@misc{springborn2013impression,
  author = {Springborn, Kevin and Barford, Paul},
  title = {{Impression Fraud in On-line Advertising via Pay-Per-View Networks}},
  year = {2013}
}

@misc{starov2018dashboard,
  author = {Starov, Oleksii and others},
  title = {{Betrayed by Your Dashboard: Discovering Malicious Campaigns via Web Analytics}},
  year = {2018}
}

@misc{stonegross2011adexchanges,
  author = {Stone-Gross, Brett and others},
  title = {{Understanding Fraudulent Activities in Online Ad Exchanges}},
  year = {2011}
}

@misc{thomas2015adinjection,
  author = {Thomas, Kurt and others},
  title = {{Ad Injection at Scale: Assessing Deceptive Advertisement Modifications}},
  year = {2015}
}

@misc{urban2020gdpr,
  author = {Urban, Tobias and others},
  title = {{Measuring the Impact of the GDPR on Data Sharing in Ad Networks}},
  year = {2020}
}

@misc{usc2013adtransparency,
  author = {Taplin, Jonathan},
  title = {{USC Annenberg Lab Ad Transparency Report: January}},
  year = {2013}
}

@misc{vekaria2024darkinventory,
  author = {Vekaria, Yash and Nithyanand, Rishab and Shafiq, Zubair},
  title = {{The Inventory Is Dark and Full of Misinformation: Understanding Ad Inventory Pooling in the Ad-Tech Supply Chain}},
  year = {2024}
}

@inproceedings{vekaria2024notifications,
  author    = {Vekaria, Yash and Nithyanand, Rishab and Shafiq, Zubair},
  title     = {Towards Multi-Stakeholder Vulnerability Notifications in the Ad-Tech Supply Chain},
  booktitle = {2026 IEEE 11th European Symposium on Security and Privacy (EuroS\&P)},
  pages     = {874--891},
  publisher = {IEEE},
  year      = {2026},
  doi       = {10.1109/EuroSP68448.2026.00061}
}

@misc{venkatadri2018facebook,
  author = {Venkatadri, Giridhari and others},
  title = {{Privacy Risks with Facebook PII-Based Targeting: Auditing a Data Broker Advertising Interface}},
  year = {2018}
}

@misc{vines2017adint,
  author = {Vines, Paul and Roesner, Franziska and Kohno, Tadayoshi},
  title = {{Exploring ADINT: Using Ad Targeting for Surveillance on a Budget}},
  year = {2017}
}

@misc{yuan2013rtb,
  author = {Yuan, Shuai and Wang, Jun and Zhao, Xiaoxue},
  title = {{Real-Time Bidding for Online Advertising: Measurement and Analysis}},
  year = {2013}
}

@misc{zarras2014malvertising,
  author = {Zarras, Apostolis and others},
  title = {{The Dark Alleys of Madison Avenue: Understanding Malicious Advertisements}},
  year = {2014}
}

@misc{zeng2022targetingbids,
  author = {Zeng, Eric and McAmis, Rachel and Kohno, Tadayoshi and Roesner, Franziska},
  title = {{What Factors Affect Targeting and Bids in Online Advertising? A Field Measurement Study}},
  year = {2022}
}

@misc{zhong2023addax,
  author = {Zhong, Ke and Ma, Yiping and Mao, Yifeng and Angel, Sebastian},
  title = {{Addax: A Fast, Private, and Accountable Ad Exchange Infrastructure}},
  year = {2023}
}

\end{document}